# Disruption Avoidance via RF Current Condensation in Magnetic Islands Produced by Off-Normal Events

A.H. Reiman[1], N. Bertelli[1], P.T. Bonoli[2], N.J. Fisch[1], S.J. Frank[2], S. Jin[1], R. Nies[1], E. Rodriguez[1]

[1]Princeton Plasma Physics Laboratory, Princeton, New Jersey 08544, United States of America
[2]Massachusetts Institute of Technology: Plasma Science and Fusion Center, Cambridge, Massachusetts 02139, United States of America

## Abstract

As tokamaks are designed and built with increasing levels of stored energy in the plasma, disruptions become increasingly dangerous. It has been reported that 95% of the disruptions in the Joint European Torus (JET) tokamak with the ITER-like wall are preceded by the growth of large locked islands, and these large islands are mostly produced by off-normal events other than neoclassical tearing modes. This paper discusses the use of RF current drive to stabilize large islands, focusing on nonlinear effects that appear when relatively high powers are used to stabilize large islands. An RF current condensation effect can concentrate the RF driven current near the center of the island, increasing the efficiency of the stabilization. A nonlinear shadowing effect can hinder the stabilization of islands if the aiming of the ray trajectories does not properly consider the nonlinear effects.

## I. Introduction

As tokamaks are designed and built with increasing levels of stored energy in the plasma, disruptions become increasingly dangerous. The ITER research plan says that "Operation of ITER will have to strongly focus on avoiding disruptions with a high success rate and on mitigating those in which avoidance techniques fail" [1]. Disruption avoidance will also be of critical importance for fusion reactors. This paper will discuss a nonlinear effect on RF driven currents that can greatly facilitate the stabilization of magnetic islands and the avoidance of disruptions.

There is presently an intensive research effort aimed at developing a disruption mitigation capability for ITER [2]. The disruption mitigation system being developed will inject shattered pellets into the plasma to radiate away the plasma energy over a short period of time when a disruption is believed to be imminent. This will avoid catastrophic damage to the device, but the intense radiation associated with each such mitigated disruption will cause some damage to the first wall, and ITER will be able to tolerate a limited number of such mitigated disruptions to keep the cumulative damage to the first wall at an acceptable level [3]. For each disruption, there will also be some risk that mitigation is not successful [3]. These considerations motivate the desire to avoid the need for mitigation to the extent possible in ITER. It will similarly be desirable to minimize the need for mitigation in fusion reactors.

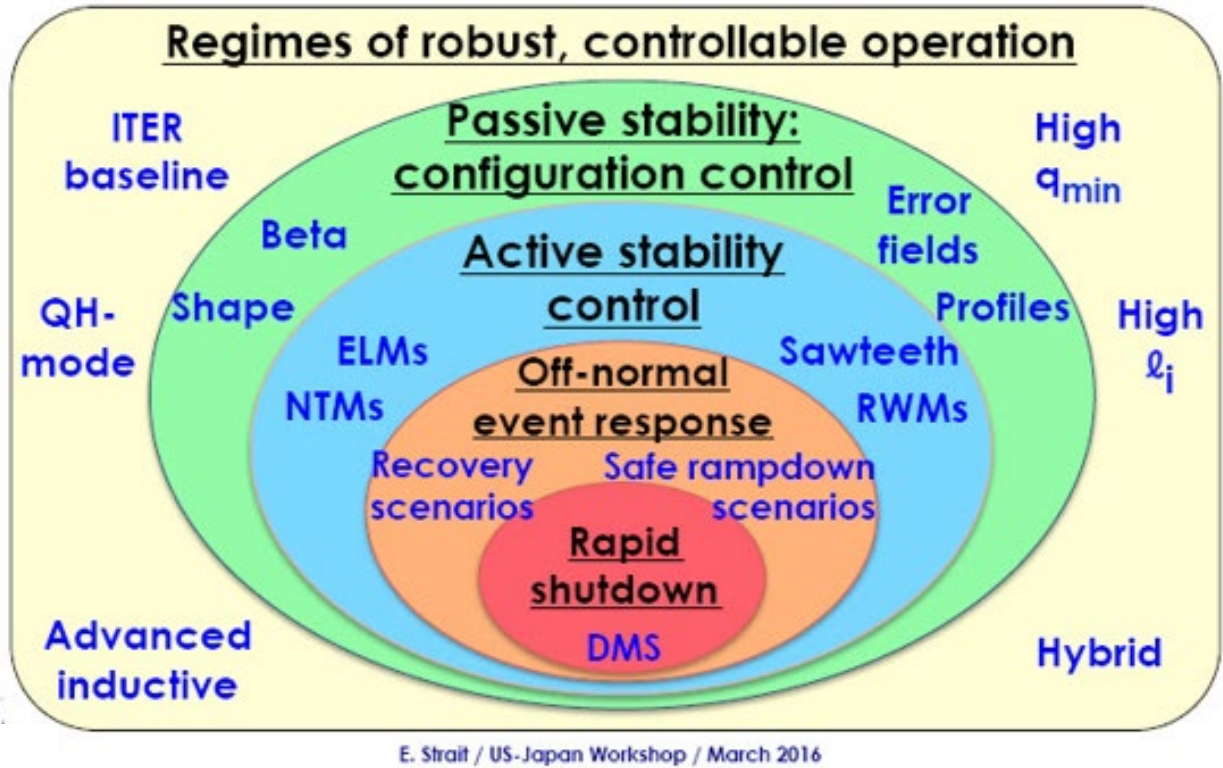

Fig. 1. *Envisioned multiple layers of defense against disruptions in ITER. Figure courtesy of E. Strait.*

It is recognized that a reliable defense against disruptions in ITER will require a multi-layer approach. Fig. 1 shows an envisioned set of defensive layers, with disruption mitigation as a last resort. This paper will be concerned with the use of RF stabilization of magnetic islands for disruption avoidance, and there are two layers in Fig. 1 that are of relevance in that regard. The routine stabilization of small islands produced by neoclassical tearing modes in ITER using electron cyclotron current drive (ECCD) can be considered to be part of the active stability control layer. In this paper we are going to be interested in the next higher layer. We will be interested in the use of ECCD to stabilize large islands produced by off-normal events. With the significant amount of EC power being installed in ITER, interest in RF stabilization of islands will focus primarily on ECCD in the near term, but stabilization by lower hybrid current drive (LHCD) will also be of potential interest over the longer term.

It has been reported that 95% of the disruptions in the Joint European Torus (JET) tokamak with the ITER-like wall are preceded by the growth of large locked islands [4]. These large islands are mostly produced by off-normal events other than neoclassical tearing modes [5]. A statistical analysis of 250 disruptions on JET found a distinct locked mode amplitude at which the plasma disrupted, and further analysis of this data concluded that that locked mode amplitude corresponded to an island width of about 30% of the minor radius [6]. These observations suggest that islands are playing a key role in triggering the disruptions, and, as will be discussed in Section II.B, they suggest that there is time to stabilize them. The results further suggest the question whether we can reduce the need for mitigation by using RF current drive to suppress the large islands produced by off-normal events. It has been demonstrated that locked islands can be stabilized by RF driven currents if the error field compensation coils are used to control the phase at which the island locks [7]. This can be accomplished by slightly overcompensating for the ambient field error.

ECCD island stabilization studies for ITER have largely focused on the stabilization of small islands produced by neoclassical tearing modes (NTMs) using as little power as possible to minimize the impact on Q. (See, e.g. Ref. [8] and references therein.) When large islands appear and cause significant deterioration of confinement and threaten to trigger disruptions in ITER (or in tokamak reactors), it will be desirable to use the full amount of available RF power to stabilize them if necessary. When relatively large amounts of power are used to stabilize large islands, nonlinear effects appear that have only recently been recognized [9]. The nonlinearities can lead to an RF current condensation effect that can be used to facilitate the stabilization of large islands, and to allow the stabilization of larger islands than would otherwise be possible. The nonlinearities can also lead to a nonlinear shadowing effect that can hinder the RF stabilization of islands if the aiming of the ray trajectories does not consider the nonlinear effects [10]. These nonlinear effects, and their implications for RF stabilization of islands, will be the subject of this paper.

Section II of the paper will provide some background information on the large islands produced by off-normal events. Section III will discuss the RF current condensation effect and its implications for the stabilization of large magnetic islands. One facet of the condensation effect is the appearance of a bifurcation in the solution of the nonlinear steady state thermal diffusion equation in the magnetic island. Section IV will discuss the saturation of the temperature when the power deposition exceeds the bifurcation threshold. A hysteresis effect associated with the bifurcation will be discussed. A picture of the nonlinear shadowing effect will also emerge from this discussion. Section V will discuss the implications of RF current condensation for the stabilization of magnetic islands using lower hybrid current drive (LHCD). Although LHCD is generally associated with relatively broad deposition profiles, the condensation effect can localize the deposition, leading to a reevaluation of its potential for island stabilization. Section VI will describe a new capability being developed for higher fidelity simulation of the nonlinear effects. It will also describe recent calculations for ITER using the new simulation capability. Finally, Section VII will present some conclusions.

## II. Background: Islands produced by off-normal events.

### A. Disruptions triggered by impurity accumulation in the plasma core.

As an example of a large island produced by an off-normal event, we consider the typical sequence of events in a JET disruption initially triggered by impurity accumulation and associated radiation in the core of the plasma. An analysis of JET shots that were run during 2011 and 2012 found that about 4.6% of the shots during this period disrupted because of such an event [5]. This was by far the major cause of disruptions in this period, causing about 15% of all the H-mode shots to disrupt. By comparison with this 4.6%, about 0.5% of the shots during this period disrupted because of neoclassical tearing modes.

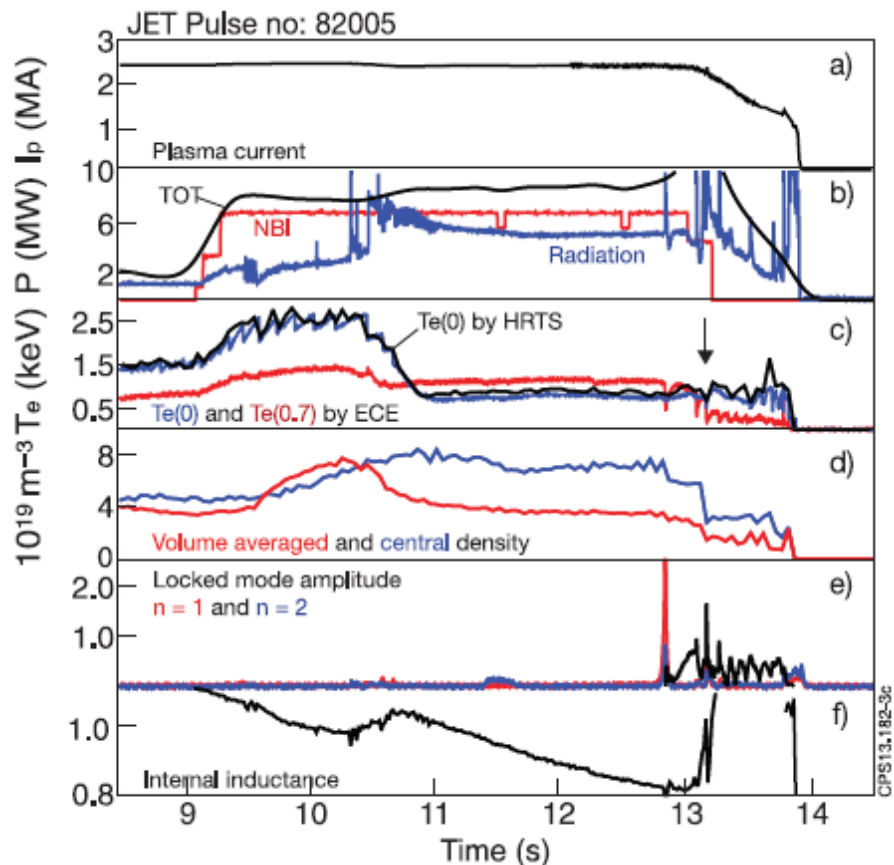

Fig. 2. *Set of time traces showing the typical sequence of events in a disruption triggered by impurity accumulation and associated radiation in the core of the plasma. Figure reproduced from Ref. [DeVries14].*

Fig. 2 displays a set of time traces showing a typical sequence of events in a disruption in JET triggered by an accumulation of impurities and associated radiation in the core of the plasma . At about 10.5 seconds there is a rapid increase in the radiation in the plasma core that cools the center of the plasma and leads to a hollow temperature profile. The hollow temperature profile increases the resistivity in the plasma core and leads to a broadening of the current profile, as indicated by the decreasing $l_i$. A tearing mode appears, believed to be driven unstable by the modification of the current profile. An island grows and quickly locks.  The locked island grows, in this case, for about 360 milliseconds before it triggers a thermal quench, but not a disruption. The locked island continues to grow for about another second before it triggers a disruption.

Despite the fact that the initial trigger for the sequence of events shown in Fig. 2 was an accumulation of impurities in the core of the plasma and the associated radiation, the direct trigger for the disruption was the growth of a magnetic island. This raises the question whether we can use ECCD to stabilize such an island and either buy enough time for the plasma to recover, or at least buy enough time for a quiescent shutdown of the plasma. More generally, looking at this sequence of events, we can ask the question: What tools do we have that can intervene on the needed time scale? That raises the question of time scales.

**B. Time scales**

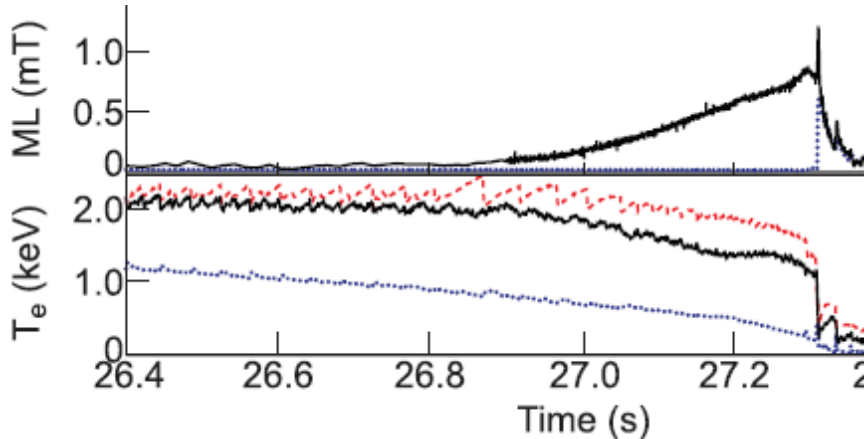

Fig. 3. *Time traces showing a disruption triggered by a born locked mode in JET. Top: The measured locked mode amplitude. Bottom: The temperature at 3 radial locations in the plasma, r/a = 0, r/a = 0.3 and r/a = 0.6, measured by the fast ECE diagnostic. Figure reproduced from Ref. [deVries16].*

Fig. 3 shows a disruption in JET triggered by a born locked mode. The locked mode begins to grow at about 26.8 seconds. It grows for about 500 milliseconds and then triggers a disruption. About 160 milliseconds before the disruption, the locked mode amplitude becomes sufficiently large to trigger a shutdown of the plasma. Although the plasma is being shut down for the last 160 ms, there is no discernible effect on the growth of the island.

Islands, rotating or locked, grow on a time scale $\Delta' a \tau_R$, where $a$ is the minor radius, $\tau_R$ is the global resistive time scale, and $\Delta'$ is the conventional resistive instability index. The resistive time scale will be much longer in ITER than it is in JET because of the higher temperature. However, the ramp-down itself occurs on a resistive time scale, and the ramp-down can itself trigger a disruption if it is done too rapidly. In contrast, RF current drive establishes a stabilizing electric field on an electron-ion collision time scale [11]. That again raises the question whether we can use RF current drive to stabilize the large islands produced by off-normal events to either buy time to recover normal operation or to at least have a safe shutdown. When large islands are produced by off-normal events in ITER, degrading performance and threatening to cause a disruption, it will be desirable to use all of the available ECCD power, if necessary, to stabilize the island. When these relatively large amounts of ECCD power are used for the purpose of stabilizing large islands, non-linear effects may come into play [9]. There is an "RF condensation" effect that can concentrate the driven current near the island center, improving the efficiency of RF current drive stabilization of islands and allowing the stabilization of larger islands than would otherwise be possible. Alternatively, there can also be a "nonlinear shadowing" effect that can prevent the RF current from being driven near the island center if the aiming of the ray trajectories does not properly consider the nonlinearity. These nonlinear effects will be the subject of the remainder of this paper.

## III. Nonlinear behavior of RF driven current in magnetic islands.

### A. The RF current condensation effect

The non-linear effects on RF driven currents in magnetic islands that will be discussed in this paper arise from the sensitivity of the RF-driven current and the RF power deposition to the temperature perturbation in the island. The conventional picture of RF current stabilization of magnetic islands, which is valid at low power, assumes that the local power deposition and electron acceleration are not affected by the presence of the island. This is shown in Fig. 4, which illustrates the conventional picture of RF current drive stabilization of a rotating island. As illustrated in Fig. 4a, the local power deposition is not affected by the presence of the island. The local electron acceleration is proportional to the local power deposition and is also not affected by the presence of the island. The accelerated electrons move rapidly along the field lines, averaging the electron acceleration over the flux surface, so that the RF driven current is a function of the flux surface. The averaging over the flux surfaces gives a higher local current density near the center of the island than near the periphery. This then produces a stabilizing resonant component of the magnetic field. It is this geometric effect that gives the stabilization of rotating islands by RF driven currents in the conventional picture.

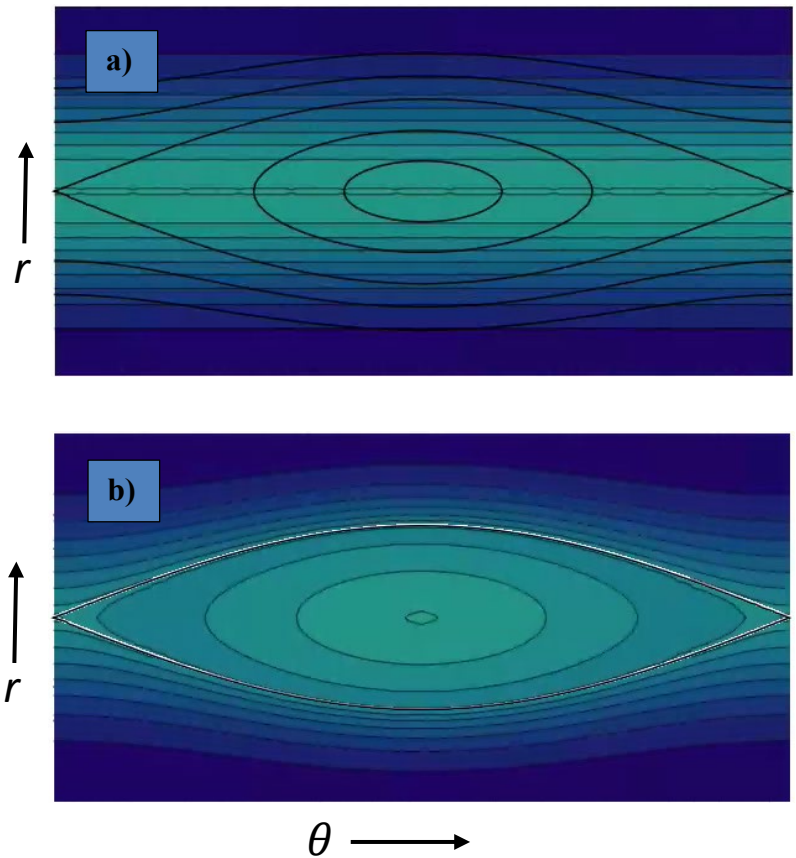

Fig. 4. *Figure illustrating the conventional picture of the RF current drive stabilization of a rotating island. a) The horizontal lines correspond to contours of the local power deposition, which is unaffected by presence of the island. b) Contours of the magnitude of the RF driven current, after taking into account the averaging over flux surfaces.*

The conventional model does not consider the sensitivity of the RF power deposition and the RF electron acceleration to the local temperature, which can become important for high RF powers and large island widths. This is illustrated in Fig. 5. (As will be discussed in Section III.C, conventional models do take into account the effect of the temperature perturbation on the Spitzer resistivity, but that is different from the effect on the RF driven current discussed here. The relation between the two effects will be discussed in that section.) The RF power deposition in the island heats the island, with the temperature peak at the island O-point. The increased temperature in the island causes the RF power deposition to increase, which further increases the temperature in the island. There is a nonlinear feedback effect, giving rise to a nonlinearly enhanced temperature in the island. The local RF electron acceleration in the island is itself sensitive to the local temperature, increasing as the temperature in the island is increased. The combination of these two effects give rise to the RF current condensation effect, which can concentrate the driven current near the center of the island. That increases the magnitude of the stabilizing resonant component of the field produced by the RF driven current. This can facilitate the stabilization of the island by the RF driven currents, allowing the stabilization of larger islands than would otherwise be possible.

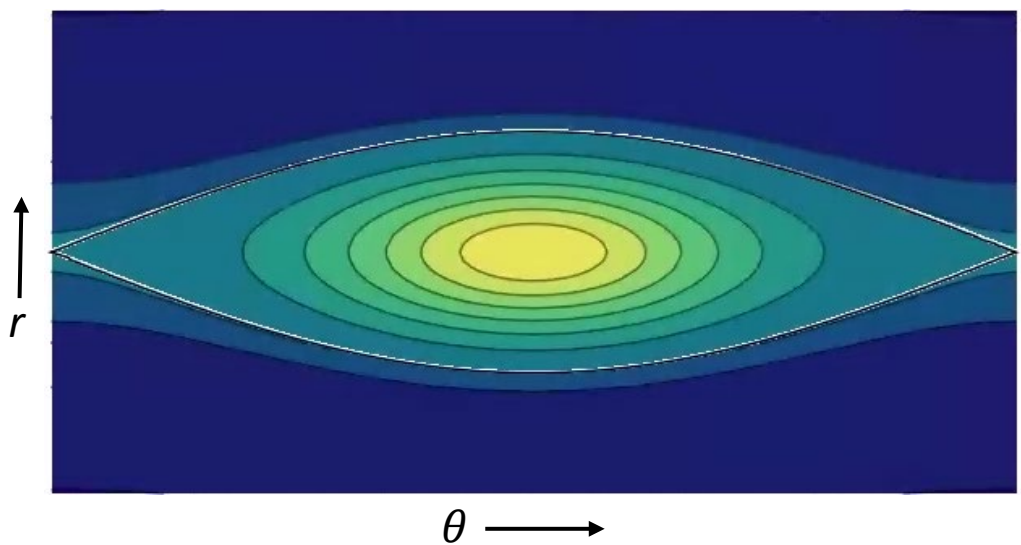

Fig. 5. *Figure illustrating the RF current condensation effect, which arises from the sensitivity of the RF power deposition and electron acceleration to the local temperature.*

**B. Sensitivity to temperature perturbation in the island.**

Electron cyclotron current drive (ECCD) [12,13] and lower hybrid current drive (LHCD) [14,15] are sensitive to the temperature because they deposit their energy on the electron tail. The further out on the tail, the more efficient is the current drive. Two-dimensional Fokker-Planck simulations for nonrelativistic resonant electrons have found that the number of such electrons is essentially determined by the lowest resonant phase velocity in the wave spectrum, both for the case of LHCD [16] and for the case of ECCD [13]. Denoting this lowest resonant phase velocity by $V_p$, the number of resonant electrons in these cases is determined by the Maxwell distribution function, and is therefore proportional to $\exp(-V_p^2/V_T^2) = \exp(-w^2)$, where $V_T$ is the thermal velocity of the electrons and we have defined $w \equiv V_p/V_T$. (This is the conventional notation used in the theory of RF driven currents. We will denote island widths by $W_i$.) Writing the temperature as a sum of an unperturbed piece plus a perturbation, $T = T_0 + \tilde{T}$, the number of resonant electrons is then proportional to

$$n_{res} \propto \exp(-w^2) = \exp(-w_0^2)\exp(w_0^2 \tilde{T}/T), \tag{1}$$

where $w_0$ is the unperturbed value of $w$.

The situation for ECCD is complicated by relativistic effects [17]. For this more general case, numerical calculations find that, if we define $w_{eff}$ such that the power deposition is proportional to $\exp(-w_{eff}^2)$, then it remains the case that $w_{eff}^2 \propto 1/T$ for a broad range of parameters of interest, so that we can write $T w_{eff}^2 \approx T_0 w_0^2$, where $w_0$ is now the unperturbed value of $w_{eff}$, $w_0 = w_{eff}(T = T_0)$ [18]. It follows that

$$\exp(-w_{eff}^2) \approx \exp(-w_0^2)\exp(w_0^2 \tilde{T}/T). \tag{2}$$

Numerical calculations of the sensitivity of the RF power deposition to temperature will be discussed for electron cyclotron waves in Section VI, and for lower hybrid waves in Section V.A. Typically, $w_0^2$ is greater than or approximately equal to 4 for ECCD and is larger for LHCD. We get a significant nonlinear effect on the power deposition and driven current when $\tilde{T}/T_0 \approx 0.5/w_0^2$, which is an experimentally relevant regime. There are two pieces to the non-linear effect. There is a nonlinear increase in the power deposition, which feeds back on itself to give a nonlinearly enhanced temperature perturbation, and there is a nonlinear increase in the driven current. We will consider each of these in turn, turning first to a consideration of the nonlinear effect on the driven current.

**C. Effect of the temperature perturbation in the island on the current density.**

The temperature perturbation in the island affects both the ohmic and the RF-driven currents in the island. The stabilization of magnetic islands through the effect of the heating of the islands on the ohmic current profile has been extensively studied, and it is believed to have provided a significant stabilizing effect in a number of experiments [19,20,21,22,23]. The temperature perturbation modifies the Spitzer resistivity, giving a fractional change in the ohmic current density $\Delta J_{Sp} / J_{Sp} = \Delta \sigma_{Sp} / \sigma_{Sp} \approx (3/2)\tilde{T}/T_0$ for $\tilde{T} \ll T_0$, where $J_{Sp}$ is the ohmic current and $\sigma_{Sp}$ is the Spitzer conductivity. This increases the relative ohmic current density near the island O-point, producing a stabilizing resonant component of the field.

By comparison with the change in the ohmic current density, the change in the RF driven current density is given by $\Delta J_{RF} / J_{RF} \approx \exp(w_0^2 \tilde{T}/T_0) - 1 > w_0^2 \tilde{T}/T_0$. Even when the fractional change in the temperature perturbation is quite small and we are in the linear regime, the coefficient in front of the fractional temperature perturbation is much larger for RF-driven currents than it is for ohmic currents. When we enter the nonlinear regime, the fractional change in the RF driven current is an exponential function of the temperature perturbation as long as $\tilde{T} \ll T_0$. When the bootstrap current density becomes comparable to the ohmic current density, the RF driven current density required to stabilize the island becomes comparable to the ohmic current density, and the effect of the temperature perturbation on the RF-driven current density becomes much more important than the effect on the ohmic current density.

The effect of the temperature perturbation on the RF driven current, and on the stabilization of the island, has been recognized for some time [11], although the implications for the stabilization of islands appear not to have been further studied since the early 1980's. More recently, it has been recognized that the effect of the temperature perturbation on the power deposition also becomes important in this context, and that the combination of the two can lead to the RF current condensation effect [9]. We turn next to a discussion of the nonlinear enhancement of the temperature perturbation in the island that arises from the sensitivity of the power deposition to the temperature perturbation.

**D. Nonlinear enhancement of the temperature perturbation in the island**

The rate of growth of the island is generally slow relative to the transport time scale in the island, so we can consider the steady-state thermal diffusion equation,

$$\nabla \cdot (n\mathbf{\kappa} \cdot \nabla T) = -P, \tag{3}$$

where $n$ is the density, $\mathbf{\kappa}$ is the thermal diffusivity tensor and $P$ is the local power deposition. The component of the thermal diffusivity parallel to the magnetic field is large relative to that perpendicular to the magnetic field, and we assume that the temperature is constant on the flux surfaces both outside and inside the island. (We assume that the island is not sufficiently small to violate this assumption [24].) For each flux surface that encircles the magnetic axis, we can integrate Eq. (3) over the volume enclosed by the flux surface to find that the temperature gradient on the flux surface is determined by the total power deposited between that flux surface and the magnetic axis and by the local thermal diffusivity. It is convenient to consider the case where the RF power is initially deposited on flux surfaces between the island and the magnetic axis and is subsequently redirected to the island interior. In that case the temperature on the island separatrix is unperturbed. The temperature in the island is initially flat and is perturbed when the RF power is redirected there.

To calculate the temperature perturbation in the island, we solve a nonlinear thermal diffusion equation there. To obtain a diffusion equation in magnetic island geometry, we employ a conventional cylindrical model for the magnetic field,

$$\mathbf{B} = \nabla \psi \times \hat{z} - (kr/m) B_z \hat{\theta} + B_z \hat{z}, \tag{4}$$

where we can expand $\psi$ about the rational surface as

$$\psi = \psi_0''(r-r_s)^2/2 - \epsilon \cos(m\zeta), \tag{5}$$

$\zeta = \theta - kz/m$, and $\epsilon$ is a constant (the "constant-psi approximation"). We define

$$\rho^2 = \psi/2\epsilon + 1/2, \tag{6}$$

and we transform our angular coordinate to $\eta = \arcsin\left[\sin(\zeta/2)/\rho\right]$. The temperature is a function only of $\rho$, $T = T(\rho)$. Defining an operator

$$\langle f \rangle(\rho) \equiv \int_{-\pi/2}^{\pi/2} J f(\rho,\eta) d\eta,$$

where $J = (\nabla\rho \cdot \nabla\eta \times \nabla\phi)^{-1}$, we can write

$$\frac{d}{d\rho}\left(n\kappa_\perp \langle |\nabla\rho|^2 \rangle \frac{d}{d\rho} T(\rho)\right) = \frac{1}{2\pi} P_{RF}(\rho) \frac{dV}{d\rho}$$

as our diffusion equation, where $V(\rho)$ is the volume inside the corresponding flux surface and $\kappa_\perp$ is the cross-field thermal conductivity.

When $\tilde{T}/T_0$ is small but $w_0^2 \tilde{T}/T_0$ is not small, the dependence of the RF power deposition, $P_{RF}$, on $\tilde{T}$ comes in entirely through the exponential dependence of the number of resonant electrons on $\tilde{T}$, $P_{RF} = P_{RF0} \exp(w_0^2 \tilde{T}/T_0)$, where $P_{RF0}$ is the power deposition in the absence of the perturbation. Taking $n$ and $\kappa_\perp$ to be constant in the island, we define a normalized fractional temperature perturbation $u \equiv w_0^2 \tilde{T}/T_0$, and a normalized power $P_0 \equiv W_i^2 w_0^2 P_{RF0}/(4 n \kappa_\perp T_0)$, where $W_i$ is the island width. After some algebra, and after discarding a term small in $W_i/R$, where $R$ is the major radius, the diffusion equation takes the form

$$\frac{d}{d\rho}\left(\frac{1}{\rho}\left[E(\rho)-(1-\rho^2)K(\rho)\right]\frac{d}{d\rho}u(\rho)\right) = P_0\rho K(\rho)\exp u, \quad (7)$$

where $K(k) \equiv \int_0^{\pi/2}(1-k^2\sin^2\chi)^{-1/2}d\chi$ is the complete elliptic integral of the first kind, and $E(k) \equiv \int_0^{\pi/2}(1-k^2\sin^2\chi)^{1/2}d\chi$ is the complete elliptic integral of the second kind. We solve the equation with the boundary conditions $u = 0$ at the separatrix and $du/d\rho = 0$ at the O-point. (The boundary condition at the separatrix has been discussed above.) Eq. (7) can be roughly approximated by using a slab model of the island interior, which gives a simplified equation that can be solved analytically when $P_0$ is a constant,

$$d^2u/dr^2 = -P_0\exp(u). \quad (8)$$

We solve Eqs (7) and (8) with $P_0$ taken to be a constant in the island. Fig. 6 shows the solutions for the normalized temperature perturbation at the O-point, $u(0)$, as a function of $P_0$ for Eq. (7), for Eq. (8), and for a linearized version of Eq. (7) where the $e^u$ factor has been neglected. It can be seen in Fig. 6 that the solutions to the nonlinear equations display a bifurcation, with the bifurcation threshold in magnetic island geometry corresponding approximately to $P_0 = 1.02$.

We can understand the bifurcated solution to the steady state diffusion equation if we consider the behavior of the solutions to the time-dependent diffusion equation. Consider first the situation when $P_0$ is below the bifurcation threshold. Initially, when the temperature in the island is flat, the diffusive, second derivative term in the diffusion equation vanishes. The power deposition term dominates initially, at low temperature, and the temperature increases. The second derivative term increases with increasing temperature, until it balances the power deposition at the lower root of the steady-state diffusion equation. A perturbation to a higher temperature gives a further increase in the second derivative, so that the lower root is stable. At sufficiently high temperature, the exponential begins to dominate, and the power deposition term increases more rapidly than the diffusive term with increasing temperature, until the two terms again balance at the second root. The power deposition term continues to increase more rapidly than the diffusive term with increasing temperature, so that the second root is unstable. The temperature then continues to increase until limited by effects not included in Eqs. (7) and (8), giving a third, stable solution branch not shown in the figure. Those effects will be the subject of Section IV. The two lower solution branches merge at the bifurcation point. Above the bifurcation point the increase of the power deposition with temperature begins to dominate before a balance with the diffusive term is reached, and the temperature rises until the uppermost solution branch is reached.

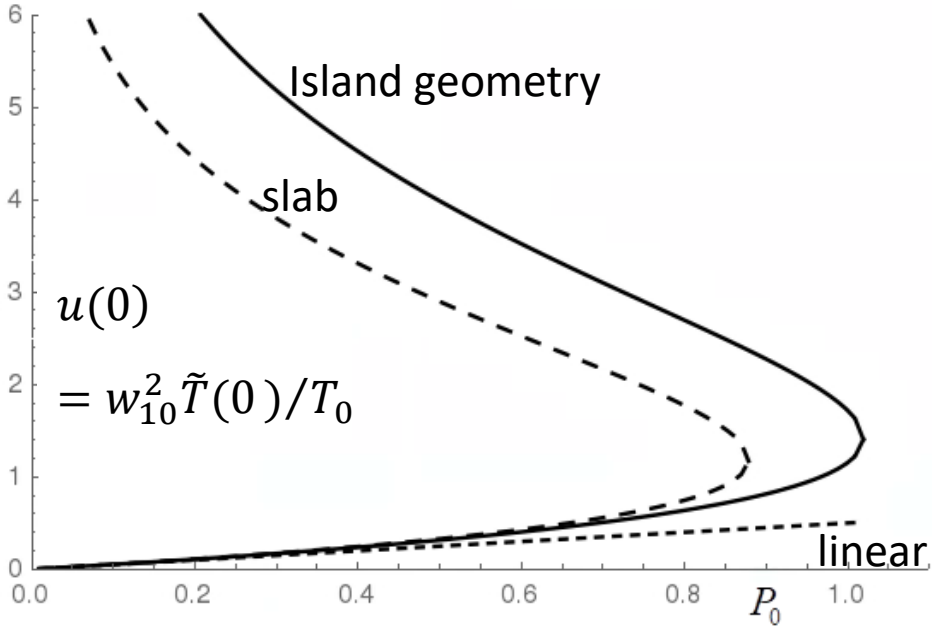

Fig. 6. *Solutions of the thermal diffusion equations in the magnetic island for an initially uniform deposition profile. The normalized temperature perturbation at the O-point is shown as a function of the normalized power for the nonlinear equations in slab and island geometry, and for the linearized equation in island geometry. Reproduced from Ref. [9].*

Before turning to the subject of the saturation of the temperature and the appearance of a third branch of the solutions to the nonlinear thermal diffusion equations, we consider the consequences of our findings thus far for the stabilization of magnetic islands.

### E. Effect of RF condensation on stabilization efficiency

As discussed in Section III.C, the exponential dependence of the RF current drive on the temperature perturbation in the magnetic island leads to a concentration of the RF driven current near the island O-point. As discussed in Section III.D, the exponential dependence of the power deposition in the island on the temperature perturbation gives rise to a nonlinearly enhanced temperature perturbation. The combination of those two effects gives rise to the RF current condensation effect. (The nonlinear enhancement of the temperature perturbation also increases the magnitude of the ohmic stabilization effect, and that may be important if the RF driven current density is small relative to the ohmic current density. As discussed in Section III.C, the modification of the RF driven current density profile is much larger when the current densities are comparable.) In this section we look at the consequences of the RF condensation effect for the efficiency of stabilization of the magnetic island. For that purpose, we adapt a simple model introduced in Ref. [20] that assumes a flat power deposition in the island. We take the deposition profile to be initially flat, and we allow it to be modified by the temperature perturbation. For the measure of efficiency, we adopt a widely used measure that was introduced in Ref. [20], the ratio of the resonant Fourier component of the current to the total RF driven current:

$$\eta_{stab} = \int_{-\infty}^{\infty} dx \oint d\zeta J_{RF}(x,\zeta) \cos(m\zeta) \Big/ \int_{-\infty}^{\infty} dx \oint d\zeta J_{RF}(x,\zeta), \qquad (9)$$

where $x$ is the radial coordinate and $\zeta$ is the helical coordinate. (The quantity $\Delta'$ in the modified Rutherford equation is proportional to the resonant component of the current.)

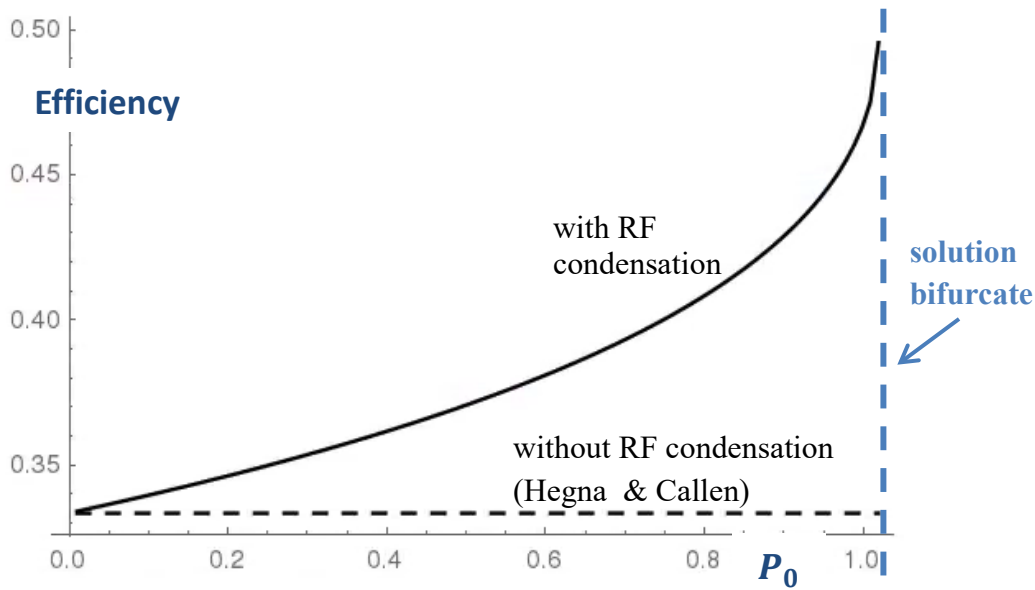

Fig. 7. Efficiency vs $P_0$ with and without the condensation effect.

Fig. 7 shows the calculated efficiency of the RF current stabilization as a function of the normalized power with and without the condensation effect. Without the effect, the efficiency is independent of the power, and we recover the result of Ref. [20]. When we include the dependence on the temperature, the RF driven current becomes increasingly concentrated near the center of the island with increasing power, and the efficiency is rising rapidly as we approach the bifurcation point.

## IV. Saturation of the temperature

Thus far, the only nonlinear effect that we have included in the calculation of the temperature is the dependence of the power deposition on the temperature perturbation in the island. Solving the resulting nonlinear, steady-state thermal diffusion equations, Eqs. (7) and (8), we have found that the solutions display a bifurcation. There is no solution when the power deposition exceeds the bifurcation threshold. The temperature rises until we encounter an additional nonlinear effect that we have not yet included in the equations. In this section, we discuss two effects that can cause the temperature to saturate: the depletion of the energy in the wave [10], and the dependence of the thermal diffusion coefficient on the temperature gradient when a microinstability threshold is encountered [25].

### A. Depletion of the energy in the wave and the hysteresis effect

We use a slab model of the island interior to study the effect of the depletion of the energy in the RF wave on the RF current condensation effect. We saw in Section III.D that the simplified equation that we obtained with this model reproduced the qualitative features of the more exact

414solution in magnetic island geometry, and it provided a rough quantitative approximation to this more exact solution.

The depletion of the energy in the wave at any given point is proportional to the energy remaining in the wave at that point. If we neglect, for the moment, the dependence of the power deposition on the temperature perturbation, and we let $\bar{V}(l)$ represent the energy density in the wave as a function of the distance along the ray trajectory, $l$, we get $d\bar{V}/dl = -f(l)\bar{V}$, where $f(l)$ is a function of position that determines the deposition profile. If we include the dependence of the power deposition on the temperature perturbation, we get $d\bar{V}/dl = -f(l)\exp(u)\bar{V}$, where $u$ is the normalized fractional temperature perturbation defined earlier. The local power deposition is $P_{RF} = -d\bar{V}/dl$. We will simplify here by taking $f$ to be a constant in the island, corresponding to an initially broad power deposition profile. A more general set of cases is considered in Ref. [10]. It will be convenient to define a dimensionless parameter $\alpha_0$ such that $f = 2\alpha_0/W_i$. We define $x$ to be a normalized distance from the rational surface such that $x \equiv (l-l_s)2/W_i$, where $l_s$ is the location of the rational surface. This places the island edges at $x = \pm 1$.

In substituting the power deposition into the diffusion equation, we now need to be careful to take into account the island topology. In the slab model of the island interior, $x = 0$ corresponds to the center of the island and $\pm x$ correspond to the same flux surface in the island interior. This was not an issue for the slab model earlier because the power deposition was symmetric about $x = 0$. Defining a normalized energy density as $V \equiv W_i w_0^2 \bar{V}/(n\kappa_\perp T_0)$, we now get the pair of equations

$$u'' = [V'(x) + V'(-x)]/2, \qquad (10)$$

$$V'(x) = -\alpha_0 e^{u(x)} V(x), \qquad (11)$$

where the prime indicates a derivative with respect to $x$. Note that $u(x)$ is an even function of $x$, and Eq. (10) is solved on the domain $0 \le x \le 1$, while Eq. (11) must be solved on the domain $-1 \le x \le 1$. To solve the coupled equations it is convenient to separate $V(x)$ into its odd and even parts, giving three coupled equations that can be solved on the domain $0 \le x \le 1$. The equations are solved subject to the boundary conditions $u(1) = 0$, and $u'(0) = 0$. We must also specify the value of $V(-1)$. In solving Eqs. (7) and (8), there was a single free parameter that needed to be specified, $P_0$. Now there are two free parameters that need to be specified, $\alpha_0$ and $V(-1)$.

The analysis of Sections III.D and III.E adopted a model that assumed that the power deposition in the island was uniform when the temperature perturbation was small, with the magnitude of the total power deposition in the island at low temperature parameterized by the normalized power $P_0$. The power deposition is no longer uniform when we include depletion of the wave energy, with the wave amplitude decreasing as the energy is depleted. The depletion is small, however, in the limit where $\alpha_0$ is small. In that limit, $V$ is a constant, and $P_0 = \alpha_0 V$. We saw that the effect of the nonlinear dependence of the power deposition on the temperature was relatively small for $P_0 < 0.5$, corresponding to $\alpha_0 V < 0.5$.

Eqs. (10) and (11) can be solved analytically. The solution is

$$u(x) = 2\ln\gamma - \ln\left[\sqrt{(\lambda+1)^2 - \gamma^2}\,\cosh(\alpha_0\gamma x) + \lambda + 1\right], \qquad (12)$$

where the parameters $\lambda$ and $\gamma$ are determined by

$$\left(\gamma^2 - 1 - \lambda\right)^2 = \left[(\lambda+1)^2 - \gamma^2\right]\cosh^2(\alpha_0\gamma), \qquad (13a)$$

$$\gamma^2 = 2\lambda + 1 + \left[V(-\alpha_0) - \lambda\right]^2. \qquad (13b)$$

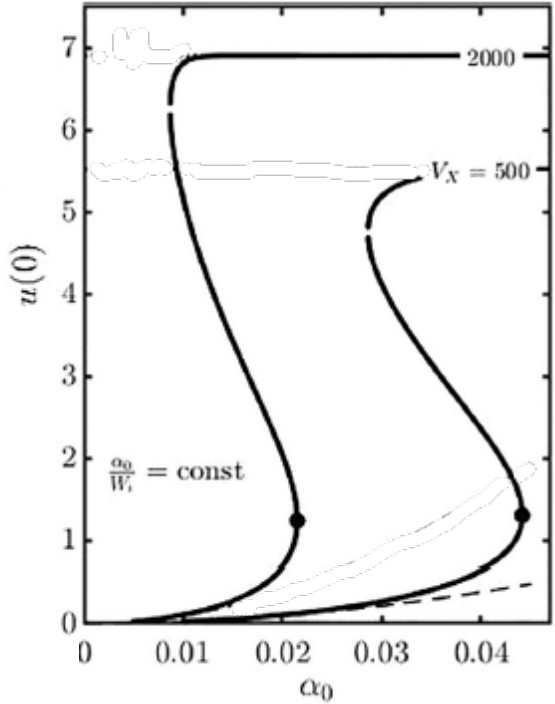

Fig. 8. *Normalized temperature perturbation at the O-point as a function of the parameter $\alpha_0$ for two different values of $V_X \equiv V(-1)/\alpha_0$. Figure reproduced from Ref. [10].*

We are particularly interested in knowing what happens when the island is stabilized and $W_i$ shrinks at fixed values of the incoming energy density, $\bar{V}(-W_i) = n\kappa_\perp T_0 V(-1)/(W_i w_0^2)$, and deposition coefficient $f = 2\alpha_0/W_i$. For that purpose we define $V_X \equiv V(-1)/\alpha_0$, and we solve for $u(0)$ as a function of $\alpha_0$ with $V_X$ fixed and $W_i \propto \alpha_0$. The results are shown in Fig. 8 for two different values of $V_X$. Comparing with Fig. 6, we see that the solution to the nonlinear thermal diffusion equation in the island has now picked up a third branch. On the lowest branch, the depletion is small enough that we can roughly approximate $V \approx V(-1)$, and the power deposition in the island can be approximated as uniform, with $P_0 \approx \alpha_0^2 V_X$. This would predict a bifurcation point at $\alpha_0^2 V_X \approx 1$, which is consistent with the results shown in the figure. Now when the power deposition exceeds the bifurcation threshold in the island the temperature rises until there is strong depletion of the energy in the island, and the temperature rise saturates on the third branch. There is a hysteresis effect, with the temperature remaining on the third branch for some time if $\alpha_0$ decreases. When the temperature rises abruptly to the third branch, the RF driven current becomes concentrated near the center of the island, and the stabilization efficiency increases. If the island is now stabilized, the island width decreases, and $\alpha_0$ correspondingly decreases. The solution remains on the upper branch for some time as the island width decreases.

Another effect that can be caused by the depletion of the energy in the wave is "shadowing" [10]. If the conventional linear expression for the power deposition is used to determine the aiming of the ray trajectory, the increased power deposition that arises from the temperature perturbation in the island may cause the energy in the wave to be depleted before the wave reaches the center of the island.

**B. Temperature stiffness at a microinstability threshold**

Another effect that can cause the temperature perturbation in the island to saturate is the profile stiffness that can be encountered when a microinstability threshold is exceeded [25]. It has been found that the temperature profile in axisymmetric tokamak plasmas can be more resilient against the effects of local heating than would be deduced from power balance considerations alone [26]. The stiffness is believed to reflect the fact that the diffusion coefficient can increase rapidly with increasing temperature gradient, and it is believed that this occurs when the temperature exceeds a microinstability threshold. This then raises the question of the effect on the local thermal diffusion coefficient in a magnetic island when the temperature is flattened and well below the microinstability threshold. There is experimental [27,28,29,30] and computational [31] evidence that the diffusion coefficient in the island is much smaller than the ambient thermal diffusion coefficient in the surrounding plasma in this case. It can be expected that the diffusion coefficient will increase when the temperature gradient in the island becomes sufficiently large to encounter a microinstability threshold.

We use a conventional model for the effect of the ITG threshold on the thermal diffusion coefficient [26],

$$\kappa_\perp = \kappa_0 \left[ 1 + \frac{\kappa_s}{\kappa_0} \left( \frac{-R}{T} \frac{dT}{dr} - k_c \right) H\left( \frac{-R}{T} \frac{dT}{dr} - k_c \right) \right], \tag{14}$$

where $\kappa_0$ is the thermal diffusivity below the ITG threshold, $H$ is a Heaviside function, $k_c$ is the ITG threshold, $R$ is the major radius and $\kappa_s/\kappa_0$ is a measure of the stiffness. The diffusivity increases linearly with increasing temperature gradient above the ITG threshold. Until now, we have been taking $\kappa_\perp$ to be independent of the temperature, and we have simply merged the value of $\kappa_\perp$ into the normalization of $P_0$ and $V$. Now we define a normalized diffusivity $\tilde{\kappa} \equiv \kappa_\perp/\kappa_0$, we use $\kappa_0$ for the normalization, and we rewrite Eq. (10) as

$$\frac{d}{dx}\left(\tilde{\kappa}\frac{d}{dx}u\right) = \frac{1}{2}[V'(x)+V'(-x)]. \tag{15}$$

The thermal diffusivity is now a function of $x$ through its dependence on the local temperature gradient. We solve the equations numerically.

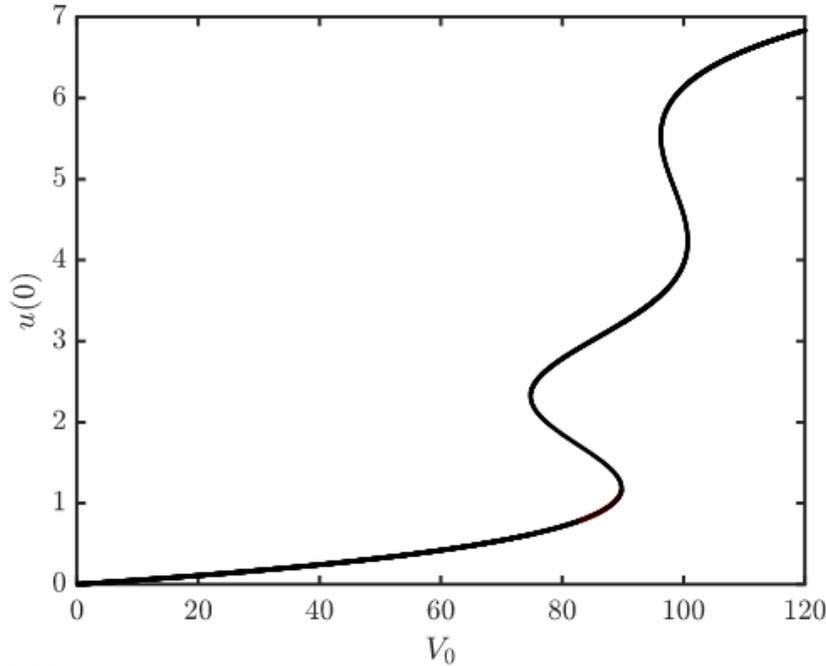

Fig. 9. Numerical solution of the equations taking into account the profile stiffness above the ITG threshold. For this set of parameters, the solution displays a double bifurcation.

Fig. 9 shows a numerical solution for $u(0)$ vs $V_0 \equiv V(-1)$ with $\alpha_0 = 0.01$, $k_c = 30$, $\kappa_s/\kappa_0 = 0.14$, and $w^2 W_i/R = 0.2$, where $V_0$ is the incoming value of $V$ at the island boundary, and we are now normalizing $V \equiv W_i w_0^2 \bar{V}/(n\kappa_0 T_0)$. When the wave power gets sufficiently large, we again encounter a bifurcation point. Above the bifurcation threshold, the temperature rises until the temperature gradient becomes sufficiently large to encounter the ITG threshold, and it continues to rise until the enhanced diffusivity becomes sufficiently large to saturate the temperature increase. If we continue to increase the power, the temperature perturbation in the island continues to increase. In this case, when the temperature gets sufficiently large we encounter a second bifurcation when the exponential dependence of the power deposition in the island becomes sufficiently large to overcome the nonlinear dependence of the thermal diffusion coefficient on the temperature. Now the temperature rises until the energy in the wave is depleted.

## V. RF current condensation with lower hybrid current drive (LHCD)

Most of the work on RF stabilization of magnetic islands has focused on the use of electron cyclotron waves for that purpose. Despite the high demonstrated efficiency of LHCD, it has generally been thought that it is not sufficiently localized for the stabilization of magnetic islands. The possibility of using RF current condensation to localize the LHCD motivates a reevaluation of the possible use of LHCD for this purpose [32]. LHCD has a particular advantage in this regard because of the relatively large value of $w^2$ at which the power tends to be absorbed. As will be discussed below, the stabilization can be further facilitated by the appropriate pulsing of the power [33]. These results raise the prospect that, in tokamaks where a significant amount of current is provided by LHCD, islands could be passively stabilized, with the LHCD automatically localizing in the island interiors, particularly as the islands become large, without requiring precise aiming of the wave power.

### A. Localization of LHCD via RF current condensation

To study the localization of LHCD via RF current condensation, simulations were performed using the General Raytracing (GENRAY) [34] and Collisional/Quasilinear 3D (CQL3D) [35,36] codes [32]. GENRAY was used to calculate the lower hybrid ray trajectories, which were then passed to the CQL3D code for calculation of the quasi-linear diffusion coefficient along the ray trajectories and the time evolution of the electron distribution function. The CQL3D time evolution was followed until the distribution function reached a steady state. For the parameters used in the calculations, the quasi-linear diffusion reaches a steady state on a time scale short compared to the thermal transport time scale in the island. The time scale separation allows the decoupling of the quasilinear diffusion calculation from the calculation of the other effects in the RF current condensation. An ITER Scenario 2 equilibrium was used for the calculations as a convenient reactor relevant plasma, with an artificial axisymmetric temperature perturbation imposed to study its effect on the localization of the LHCD. The waves were launched at a frequency of 5 GHz from a 0.5 m high waveguide grill, with a total launched power of 20 MW. High field side launch was used to provide adequate accessibility of the waves. Rays were launched at 12 different locations along the waveguide grill. At each of these locations, 20 rays were launched, with a spectral width $\Delta n_\parallel = 0.06$.

Fig. 10 shows the results of a set of calculations where the unperturbed LH power deposition was centered on the $q = 2$ surface. The LHCD launcher was located 60 degrees above the inboard mid-plane and the spectrum centered at $n_\parallel = 1.59$. Temperature perturbations up to 15% were imposed. (Experimental temperature perturbations in islands as high as 20% have been reported [21].) It can be seen in Fig. 10 that the effect of the temperature perturbation on the power deposition provides significant localization. The calculations of Ref. [32] indicate that LHCD is a potentially promising method for stabilizing magnetic islands.

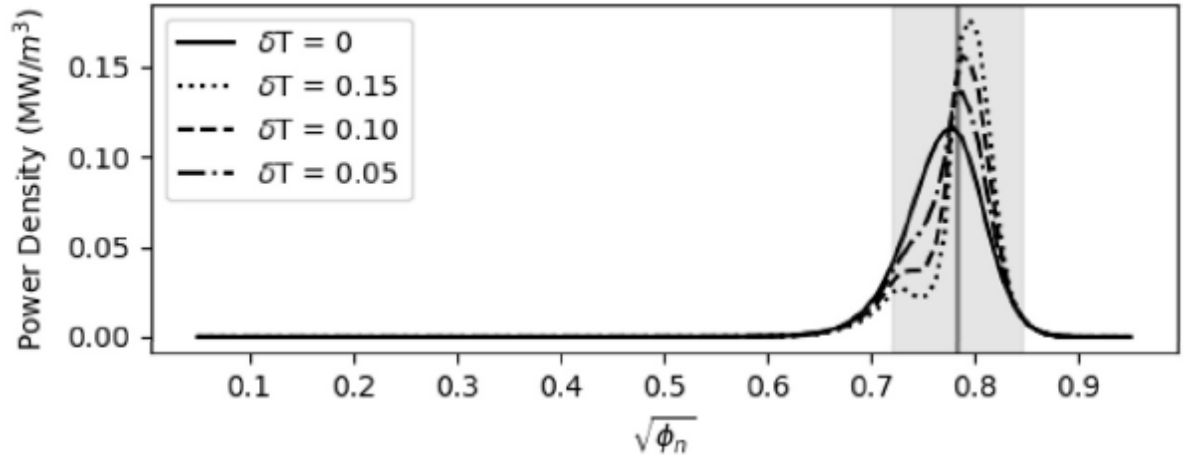

Fig. 10. *Lower hybrid wave power deposition profile at the $q = 2$ surface for four different temperature perturbation amplitudes, with the temperature perturbed in the shaded region. The radial coordinate is the square root of the normalized toroidal flux. Figure reproduced from Ref. [32].*

## B. Efficient stabilization via pulsing of the RF power

The localization of LHCD via RF current condensation offers a possible method of passively stabilizing magnetic islands in a tokamak fusion reactor. If a relatively broad deposition profile is produced in the absence of magnetic islands, the LHCD will localize wherever islands appear within that profile. The condensation effect increases in strength if an island grows wider. This approach could, potentially, dispense with the need to actively aim RF ray trajectories in order to stabilize magnetic islands when they appear. One issue that can arise is that, when the islands become large enough and the condensation effect becomes strong enough, the shadowing effect described in Section IV.A can cause the LH power to be absorbed before it reaches the center of the island. When this occurs, the stabilization efficiency becomes less optimal. This problem could be solved by appropriately adjusting the aiming of the deposition profile, but it is desirable to solve the problem without requiring active aiming of the wave trajectory. That can be accomplished by appropriate pulsing of the power [33].

The use of pulsing is motivated by the observation that, if one solves the time dependent thermal diffusion equation for an initially broad power deposition profile, the power deposition is peaked on axis early in time. Central heating initially accelerates as the absorption improves. For a sufficiently high power, or a sufficiently large island, however, there is a transition where the enhanced power deposition due to the rising island temperature causes the incoming wave to be increasingly depleted before reaching the center of the island, and the peak deposition no longer occurs in the center. If the power is now turned off, the temperature in the island relaxes back, and the optimal deposition profile can be recovered.

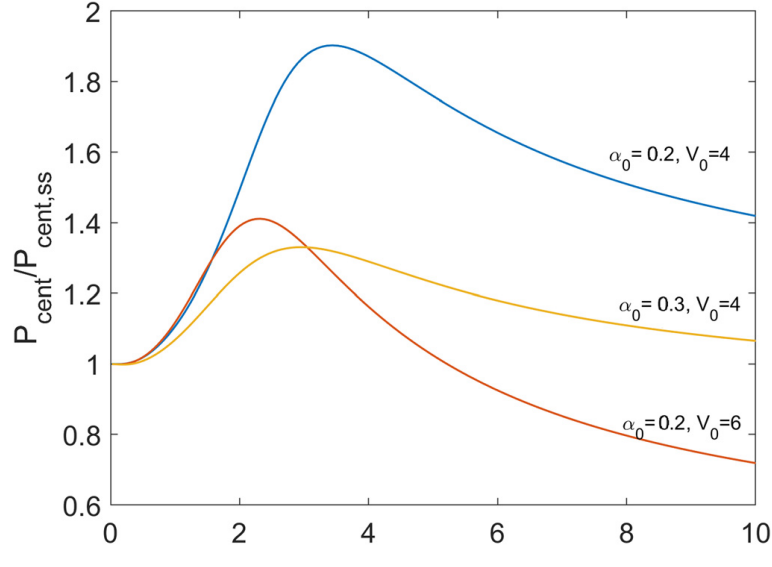

Fig. 11. *Comparison of relative $P_{cent}$ improvements for 3 representative cases performed at a duty cycle of $d = 0.25$. Figure reproduced from Ref. [33].*

The investigation of pulsing requires the solution of the time-dependent nonlinear thermal diffusion equation in the island. For that purpose, Eq. (10) is extended to include the time dependence of $u$, $\dot{u} - u'' = -[V'(x) + V'(-x)]/2$. The time is normalized to the electron thermal diffusion time scale, $t_{D,e} = 3nW_i^2/8\kappa_\perp$. We now take $V(-1) = V_0 f(t)$ for the boundary condition on Eq. (11). The time dependence of $f(t)$ is taken to be $f(t) = 1$ if $t \bmod \tau < \tau_{on}$, and $f(t) = 0$ otherwise, where $\tau$ is the period of the pulsing. The duty cycle is denoted by $d = \tau_{on}/\tau$. As a measure of the improvement obtained with pulsing, the metric $P_{cent} = \int_{-0.5}^{0.5} P dx / V_0$ is used. Fig. 11 shows a comparison of the relative improvement obtained for 3 different cases as a function of the pulsing period.

## VI. A higher fidelity simulation capability for RF current condensation

We have thus far described a series of models that provide qualitative insight into the physics of RF current condensation. In this section we turn to a discussion of a higher fidelity simulation capability that is being developed [18]. The code, OCCAMI, couples the GENRAY ray tracing code to a solution of the thermal diffusion equation in magnetic island geometry. A model nonaxisymmetric equilibrium is constructed for this purpose by starting with a numerical axisymmetric equilibrium solution and embedding the island model given by Eqs. (4) and (5) in the neighborhood of the rational surface. The perturbation of the magnetic field associated with the island is small, and it does not need to be included in the ray tracing. The calculation of the power deposition along the ray trajectory requires a specification of the density and temperature along the trajectory, and that specification must take into account the density and temperature profile in the island. This is illustrated in Fig. 12, which shows a set of ray trajectories through the island. The temperature and density are initially flattened in the island, and the temperature in the island is modified as the code iterates.

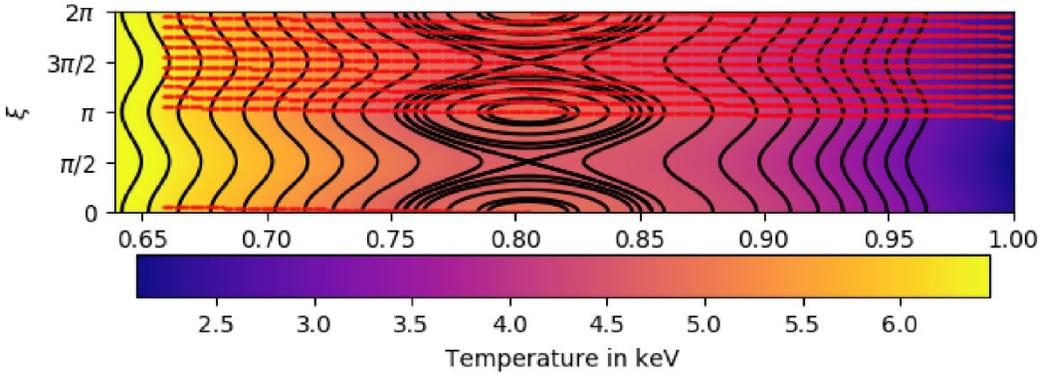

Fig. 12. A set of trajectories traversing the island region. The RF power deposition is calculated along the trajectories, including the effects of the modification of the temperature and density by the presence of the island.

Fig. 13 shows the flow of logic in the OCCAMI code. The calculation starts with initially flattened temperature and density profiles in the island. The GENRAY code is used to calculate a set of ray trajectories and the power deposition along the ray trajectories. The calculated power deposition in the island is substituted into the thermal diffusion equation for the perturbed temperature in the island, which corresponds to Eq. (7) with the $P_0 \exp(u)$ factor on the right hand side replaced by the numerically calculated power deposition. This provides an updated temperature profile in the island, which is used to recalculate the power deposition there. The diffusion equation is solved again to update the temperature profile. The code continues around this loop until the solution converges.

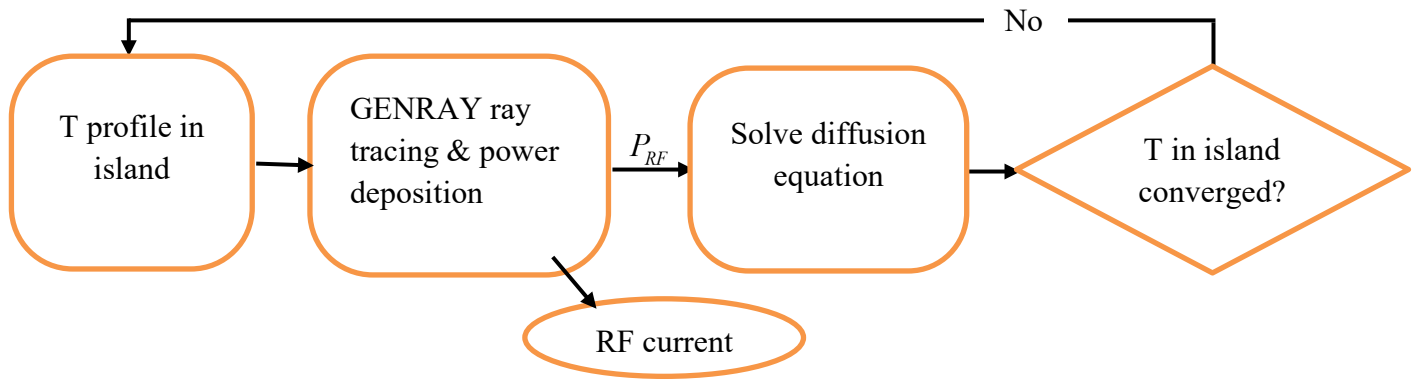

Fig. 13. Flow of logic in the OCCAMI code.

The code calculates the sensitivity of the power deposition to the temperature and expresses it in terms of an effective value of $w_0 = w_{eff}(T = T_0)$. (See discussion in Section III.B.) The calculated value of $w_0$ is a function of position, and for purposes of characterizing the impact of $w_0^2$ on the calculated results, an averaged value of $w_0^2$ is defined, $\bar{w}_0^2 \equiv -\ln\left[\langle \exp(-w_0^2)\rangle\right]$, where $\langle \ \rangle$ indicates the mean value within the island.

Fig. 14 shows the results of over 20,000 calculations using the simulation code to look at the bifurcation threshold for ITER equilibria as a function of the electron cyclotron poloidal and toroidal launch angles and launch position. An island width of 20% of the minor radius at the $q = 2$ surface was assumed. (Recall that the island width threshold for triggering a disruption in JET has been estimated to be approximately 30% [6].) For each of the calculations the EC power was increased until either a bifurcation was encountered or the maximum available power of 20 MW was reached. The figure shows the fractional temperature perturbation at the bifurcation threshold, for those calculations that did encounter a bifurcation, as a function of the parameter $\bar{w}_0^2$. The red points in the plot correspond to the ITER Scenario 2 baseline H-mode equilibrium, and the blue points represent the results for a corresponding model L-mode equilibrium. The calculations have assumed a constant diffusion coefficient in the island, and have not included the stiffness effect at the ITG threshold. The inclusion of this effect has been left for future work. Nevertheless, the calculations suggest that the bifurcation threshold will be accessible in ITER. The bifurcation threshold in each of these cases occurs at a value of $\bar{w}_0^2 \tilde{T}/T_0 > 1.6$. By comparison the nonlinear effects discussed in this paper become significant when $\bar{w}_0^2 \tilde{T}/T_0 \approx 0.5$, at a value of the temperature perturbation less than one third that at which a bifurcation is encountered. As discussed in Section III.E, the stabilization efficiency is already significantly increased at that point.

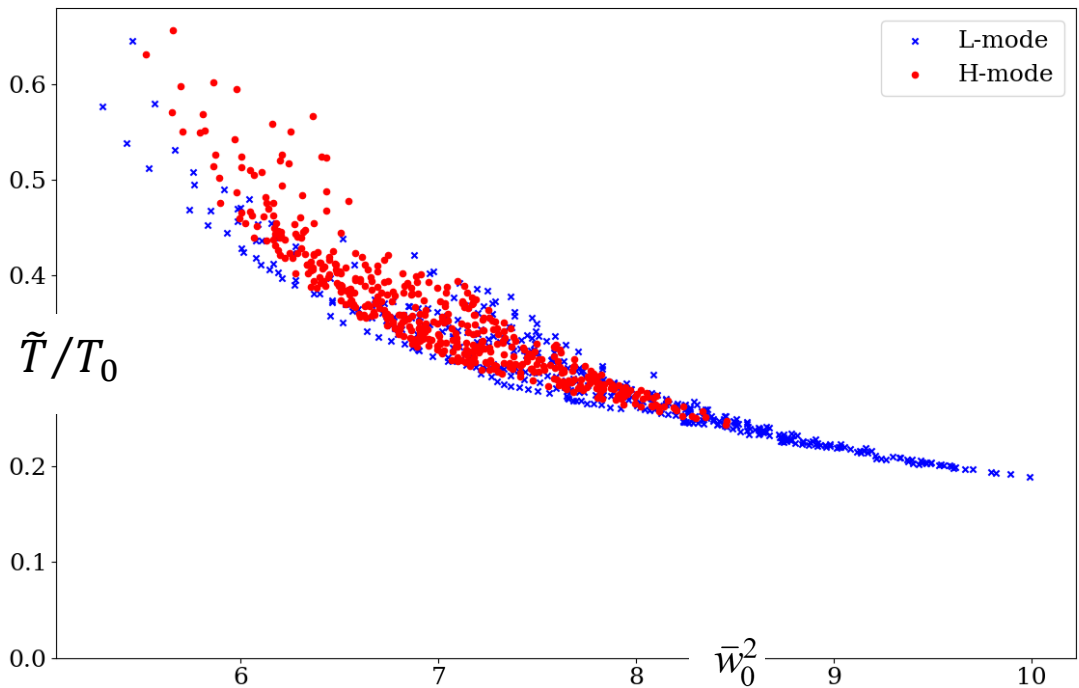

Fig. 14. *Fractional temperature perturbation at the island O-point at the bifurcation threshold for ITER-like H-mode and L-mode equilibria vs. calculated $\bar{w}_0^2$. Nonlinear effects become significant for these equilibria at a temperature perturbation $\tilde{T}/T_0$ approximately one third that at the bifurcation threshold. Figure reproduced from Ref. [18].*

For the calculations whose results are shown in Fig. 14, a range of launch positions was used. The largest values of $\bar{w}_0^2$, and therefore the lowest thresholds, were obtained for top launch. The launch position in ITER will be close to top launch.

## VII. Discussion and Conclusions

Electron cyclotron current drive (ECCD) island stabilization studies for ITER have largely focused on the stabilization of small islands produced by neoclassical tearing modes (NTMs) using as little power as possible to minimize the impact on Q. (See, e.g. Ref. [8] and references therein.) In JET, most of the large islands that cause disruptions are produced by off-normal events other than neoclassical tearing modes (NTMs) [5]. We are interested in providing a layer of protection that is specifically targeted to deal with large islands of this type.

In JET, 95% of the disruptions are preceded by the appearance of large locked islands [4]. A statistical analysis of locked mode amplitudes in JET disruptions [6], as well as an analysis of the typical sequence of events in disruptions triggered by impurity accumulation in the plasma core [5], suggest that the islands are playing a key role in directly triggering the disruptions. These islands are a target of opportunity for RF stabilization. The island width threshold for triggering a disruption in JET has been estimated to be approximately 30% [6]. The islands grow on a resistive time scale, potentially providing ample time for RF stabilization. It has been demonstrated that locked islands can be stabilized by RF current drive if the error field compensation coils are used to control the phase at which the island locks [7]. This can be accomplished by slightly overcompensating for the ambient field error.

When an island appears that is sufficiently large to cause significant deterioration of confinement and threatens to trigger a disruption in ITER or in a tokamak fusion reactor, it will be desirable to use the full available RF current drive power to stabilize the island if necessary. ITER will have 20 megawatts of ECCD power available for this purpose  For such relatively high powers and large islands, the sensitivity of the RF power deposition and of the local RF electron acceleration to the temperature perturbation in the island can lead to significant non-linear effects [9]. Both lower hybrid current drive (LHCD) and ECCD are exponentially sensitive to small perturbations of the temperature.

The exponential sensitivity of the RF power deposition in the island to the temperature perturbation there leads to a nonlinear feedback effect that gives a nonlinearly enhanced temperature perturbation. The combination of the nonlinearly enhanced temperature perturbation with the exponential sensitivity of the RF current drive to the temperature gives rise to the RF current condensation effect, which can concentrate the current in the center of the island. The condensation effect increases the stabilization efficiency of the RF driven current, and it can be used to facilitate island stabilization if the nonlinear effects are properly accounted for in the aiming of the ray trajectories [9]. Alternatively, failure to properly account for the nonlinear effects in the aiming of the ray trajectories can lead to a shadowing effect that causes the energy in the wave to be prematurely depleted, impairing stabilization [25].

When $\tilde{T} \ll T_0$, the exponential dependence of the power deposition on $\tilde{T}$ given by Eqs. (1) and (2) can be approximated as $\exp(w_0^2 \tilde{T}/T_0)$, and the steady state thermal diffusion equation in the island takes the form of Eq. (7), which can be roughly approximated by using a slab model of the island interior, Eq. (8). It is typically the case that $w_0^2 \gg 1$, so that $w_0^2 \tilde{T}/T_0$ may not be small even when $\tilde{T}/T_0$ is. The solutions to Eqs. (7) and (8) then display a bifurcation. When the power is above the bifurcation threshold there is no solution to these steady state equations, and the temperature rises until it becomes sufficiently large to bring in additional effects not included in these equations.

If $\tilde{T}/T_0$ increases to the point where it is no longer small, then we can no longer approximate $T \approx T_0$ in Eqs. (1) and (2), and the resulting expression leads to a saturation of the temperature. There are, however, two other effects that generally become important before this point. Those effects are depletion of the wave energy [10], and the profile stiffness that is encountered at a microinstability threshold [25]. Regardless of the effect causing the saturation, a third branch to the solution of the nonlinear steady-state thermal diffusion equation is introduced. There is a hysteresis effect, with the island remaining on the third branch for some time even if the stabilization and shrinking of the island width brings the normalized power back below the bifurcation threshold. The depletion of the wave energy introduces the nonlinear shadowing effect. The stiffness effect can lead to a double bifurcation as the RF power is increased, with the temperature first saturating near the microinstability threshold, and then encountering a second bifurcation point when the exponential dependence of the power deposition on the temperature perturbation becomes strong enough to overcome the nonlinear dependence of the diffusivity on the temperature above the microinstability threshold.

The RF current condensation effect motivates a reevaluation of the potential use of LHCD for the stabilization of islands. Broad LHCD deposition profiles have discouraged interest in the use of LHCD for this purpose. Ray tracing calculations using the GENRAY ray tracing code, coupled to Fokker-Planck calculations with the CQL3D code, indicate that, with the localization introduced by RF condensation, LHCD is potentially promising for stabilization [32]. The experimentally demonstrated high efficiency of lower hybrid current drive, along with the particularly strong sensitivity of LHCD to temperature perturbations arising from the high phase velocities at which it is absorbed, suggest an intriguing prospect of using LHCD for passive stabilization of islands. In this scenario, the LHCD would be automatically localized in the interior of islands as they appear and grow, with no need for active aiming of the ray trajectories. The efficiency of the stabilization can be further improved by pulsing with the appropriate frequency and duty cycle [33].

A higher fidelity simulation capability for the RF current condensation effect, the OCCAMI code is presently under development [18]. The code self-consistently couples ray tracing, power deposition and current drive calculations with the GENRAY code to the solution of the steady state thermal diffusion equation in the magnetic island. Calculations for ITER equilibria indicate that the bifurcation threshold will be accessible with the 20 MW of available ECCD power. Nonlinear effects become significant at a temperature perturbation about one third that at which the bifurcation threshold is encountered and, as discussed in Section III.E, the improvement in stabilization efficiency can already be significant at that point. If the EC toroidal launch angle is adjusted in ITER to provide efficient stabilization of large islands, the nonlinear effects described in this paper will likely become significant.

Disruptions are a critical issue for ITER and for tokamak fusion reactors. RF current drive stabilization of magnetic islands could play a major role in protecting against disruptions. An understanding of the nonlinear effects that can come into play in that context will be crucial for proper optimization of the capability. A validated predictive capability will be needed for that purpose. This will require experimental validation of the theoretical models that we are developing.

## Acknowledgments

This work was supported in part by Scientific Discovery through Advanced Computing Grant Nos. DE-SC0018090 and US DOE Grant Nos.: DE-FG02-91ER54109, DE-AC02-09CH11466, and DE-SC0016072.

The data that support the findings of this study are available from the corresponding author upon reasonable request.

## References

[1] ITER Technical Report ITR-18-003, "ITER Research Plan within the Staged Approach (Level III – Provisional Version), Sept. 17, 2018, https://www.iter.org/technical-reports.
[2] M. Lehnen, D. Campbell, D. Hu *et al*, "R&D for reliable disruption mitigation in ITER", *Preprint: 2018 IAEA Fusion Energy Conf.* Gandinaghar, India, 22–27 October 2018 https://conferences.iaea.org/event/151/papers/6095/files/4605-IAEA_2018_ITER_DMS_RD_paper_v1.4.pdf

[3] M. Lehnen, P. Aleynikov, B. Bazylev *et al*, "Plasma disruption management in ITER", Proceedings of the 26th IAEA FEC, EX/P6-39. Kyoto, Japan, 2016: Proceedings of the 26th IAEA FEC.
https://nucleus.iaea.org/sites/fusionportal/Shared%20Documents/FEC%202016/fec2016-preprints/preprint0314.pdf
[4] S. N. Gerasimov et al, *2018 IAEA Fusion Energy Conf.* Gandinaghar, India, 22–27 October 2018, IAEA-CN-258/151 (2018).
[5] P.C. de Vries, M. Baruzzo, G. M. D. Hogeweij *et al*, Phys. Plasmas **21**, 056101 (2014).
[6] P.C. de Vries, G. Pautasso, E. Nardon *et al*, Nucl. Fusion **56**, 026007 (2016).
[7] 57F. Volpe, A. Hyatt, R. La Haye *et al*, Phys. Rev. Lett. **115**, 175002 (2015).
[8] F.M. Poli *et al*, Nucl. Fusion **58**, 016007 (2018).
[9] A. H. Reiman and N. J. Fisch, Phys. Rev. Lett. **121**, 225001 (2018).
[10] E. Rodrıguez, A. H. Reiman, and N. J. Fisch, Phys. Plasmas **26**, 092511 (2019).
[11] A. H. Reiman, Phys. Fluids **26**, 1338 (1983).
[12] N. J. Fisch and A. H. Boozer, Phys. Rev. Lett. **45**, 720 (1980).
[13] C. F. F. Karney and N. J. Fisch, Nucl. Fusion **21**, 1549 (1981).
[14] N. J. Fisch, Phys. Rev. Lett. **41**, 873 (1978).
[15] N. J. Fisch, Rev. Mod. Phys. **59**, 175 (1987).
[16] C. F. F. Karney, and N. J. Fisch, Phys. Fluids **22**, 1817 (1979).
[17] I. Fidone, G. Granata, and R. L. Meyer, Phys. Fluids **25**, 2249 (1982).
[18] R. Nies, A. H. Reiman, E. Rodriguez, N. Bertelli and N. J. Fisch, Phys. Plasmas **27**, 092503 (2020).
[19] G. Kurita et al., Nucl. Fusion **34**, 1497 (1994).
[20] C. C. Hegna and J. D. Callen, Phys. Plasma **4**, 2940 (1997).
[21] E. Westerhof et al., Nucl. Fusion **47**, 85 (2007).
[22] D. De Lazzari and E. Westerhof, Nucl. Fusion **49**, 075002 (2009).
[23] P. Maget et al., Physics of Plasmas **25**, 022514 (2018).
[24] R. Fitzpatrick, Phys. Plasma **2**, 825 (1995).
[25] E. Rodrıguez, A. H. Reiman, and N. J. Fisch, Phys. Plasmas **27**, 042306 (2020).
[26] X Garbet *et al*, Plasma Phys. Control. Fusion **46**, 1351 (2004).
[27] S. Inagaki *et al*., Phys. Rev. Lett. **92**, 055002 (2004).
[28] G.W. Spakman, G.M.D. Hogeweij, R.J.E. Jaspers, *et al,* Nucl. Fusion **48**, 115005 (2008).
[29] K. Ida *et al*., Phys. Rev. Lett. **109**, 065001 (2012).
[30] L. Bardoczi *et al*., Phys. Plasma **23**, 052507 (2016).
[31] W. A. Hornsby, M. Siccinio, A. G. Peeters, *et al*, Plasma Phys. Controlled Fusion **53**, 054008 (2011).
[32] S.J. Frank, A.H. Reiman, N.J. Fisch and P.T. Bonoli, Nucl. Fusion **60**, 096027 (2020).
[33] S. Jin, N. J. Fisch, and A. H. Reiman, Phys. Plasmas **27**, 062508 (2020).
[34] A.P. Smirnov, R.W. Harvey and K. Kupfer, Bull Amer. Phys. Soc. **39**, 1626 (1994).

[35] R.W. Harvey and M.G. McCoy, 1993, The CQL3D Fokker-Planck Code, Proc. IAEA Technical Meeting on Numerical Modeling of Plasmas (Montreal, Canada, 1992) (Vienna: IAEA) (https://www.compxco.com/cql3d_manual_150122.pdf)

[36] Y.V. Petrov and R.W. Harvey, Plasma Phys. Control. Fusion **58**, 115001 (2016).